\documentclass[12pt]{article}
\usepackage{epsfig}
\usepackage{cite}
\usepackage{amsmath}
\usepackage{graphics}

\begin{document}

\setlength{\textheight}{21.5cm}
\setlength{\oddsidemargin}{0.cm}
\setlength{\evensidemargin}{0.cm}
\setlength{\topmargin}{0.cm}
\setlength{\footskip}{1cm}
\setlength{\arraycolsep}{2pt}

\renewcommand{\thefootnote}{\#\arabic{footnote}}
\setcounter{footnote}{0}

\newcommand{\gtrsim}{ \mathop{}_{\textstyle \sim}^{\textstyle >} }
\newcommand{\lesssim}{ \mathop{}_{\textstyle \sim}^{\textstyle <} }
\newcommand{\rem}[1]{{\bf #1}}
\renewcommand{\thefootnote}{\fnsymbol{footnote}}
\setcounter{footnote}{0}
\begin{titlepage}
\def\thefootnote{\fnsymbol{footnote}}

\begin{center}
\hfill
February 2008\\
\vskip .5in
\bigskip
\bigskip
{\Large \bf Multifluid Models for Cyclic Cosmology}

\vskip .45in

{\bf Irina Aref'eva}

\vskip .45in

{\it Steklov Mathematical Institute, Russian Academy of Science,\\
Gubkina st. 8, 119991, Moscow, Russia}

\vskip .45in

{\bf Paul H. Frampton and Shinya Matsuzaki}

\vskip .45in

{\it Department of Physics and Astronomy, University of North Carolina,\\
Chapel Hill, NC 27599.}

\end{center}

\vskip .4in
\begin{abstract}
Inspired by the Landau two-fluid model of superfluidity,
we consider a similar multifluid description for cosmology
where two normal fluids occur for matter and radiation
respectively. For
cyclic cosmology, two dark energy
superfluid components turn out to be insufficient but three
superfluids
can lead to a sensible five-fluid model which in a certain
limit becomes indistinguishable from the brane-world cyclic
model proposed earlier (Baum and Frampton).
Distinguishing more general five-fluid models
from brane-world
models for cyclic cosmology could be feasible with
more accurate observational data.
\end{abstract}
\end{titlepage}

\renewcommand{\thepage}{\arabic{page}}
\setcounter{page}{1}
\renewcommand{\thefootnote}{\#\arabic{footnote}}

\newpage

\noindent {\it Introduction}. ~~ In the history of physics, it is impossible to
exaggerate the fecundity of cross-fertilization between sub-disciplines. In theoretical
physics, high-energy physics and cosmology have been repeatedly informed by condensed
matter theory. One outstanding example is the idea of spontaneous symmetry breaking
introduced by Nambu\cite{Nambu} into particle theory inspired by study of the BCS
theory\cite{BCS} of superconductivity. The origin of the idea of spontaneous symmetry
breaking is in the Bogoliubov microscopic theory of superfluidity \cite{NNB}, where the
nonvanishing expectation value of the field describes the superfluid component of the
Bose liquid; this idea was exploited also in the theory of superconductivity \cite{BTS}.
Few ideas have had more impact on our understanding of both high energy physics and
cosmology.

\bigskip

Here we take our inspiration for study of cyclic cosmology\cite{BF,F,BF2}
from the Landau two-fluid
model\cite{Landau} of superfluidity, itself also applicable to
superconductivity\cite{Bardeen,BTS}; the microscopic theory of superfluidity was
developed by Bogoliubov \cite{NNB}. In this model, the energy density of superfluid
liquid helium is expressed as a sum of two terms

\begin{equation}
\rho = \rho_n + \rho_s,
\label{Landau}
\end{equation}
where $\rho_n$ is for the normal component and $\rho_s$ is for
the superfluid component.

\bigskip

At the lambda temperature, only normal fluid is present. As temperature
is decreased, more and more normal fluid is converted to
superfluid until at absolute zero the liquid helium consists
only of superfluid.
Normal fluids behave like a Newtonian fluid with viscosity and
entropy. The superfluid components have no viscosity, no entropy
and do not carry any heat.
The normal fluid and superfluid satisfy different equations of motion.

\bigskip

Similarly the different cosmological fluids will satisfy
different equations of state. One dark energy superfluid
(density $\rho_1$, equation of state $w_1$ )
is the one currently measurable. Of the others,
$\rho_2$ ($w_2$) becomes important
very close to the turnaround and $\rho_3$ ($w_3$)
is important very close to the bounce.

\bigskip

Thus, in this analogy it is natural to associate superfluids with
dark energy because it has no entropy. The normal fluids will
correspond with the matter and radiation. Thus there are $n_n=2$
normal fluids for cosmology, and we shall show that, for
cyclic cosmology, we need $n_s = 3$ dark energy superfluids,
making overall a five-fluid model.
Another approach to a dark energy fluid is in \cite{NO}.

\bigskip

\noindent {\it Friedman equation}. ~~
We write the Friedman equation as

\begin{equation}
\left( \frac{\dot{a}}{a} \right)^2
= \frac{8 \pi G}{3} \left[ \frac{(\rho_m)_0}{a(t)^3}
+ \frac{(\rho_r)_0}{a(t)^4}
+ \Sigma_{i=1}^{i=n_s} (\rho_i)_0 a(t)^{3\phi_i} \right] ,
\label{Friedman}
\end{equation}
where the equations of state for the normal fluids are
$w_m = 0$ and $w_r = +1/3$ for matter and radiation respectively.
The number of superfluids representing dark energy is $n_s$
and the first component with {\it ``$i=1$"} will be the presently
observed dark energy with $w_1 = -1 - \phi_1$ and $\phi_1 > 0$.

\bigskip

In Eq.(\ref{Friedman}), the $n_s$ terms in the summation
are analogs of the superfluid term in the Landau theory
in that they carry zero entropy. To implement cyclic
cosmology, it will be necessary that some
(actually those with {\it $i \ge 2$})
superfluid energy densities
$(\rho_i)$ be negative.

\bigskip

Let us first consider an $n_s=2$ four-fluid model with
$\phi_2  > \phi_1$, that is $w_2 = -1 - \phi_2 < w_1$.
For turnaround from expansion to contraction
at time $t=t_T$, and using $a(t_0) = 1$, we see that

\begin{equation}
(\rho_2)_0 = - (\rho_1)_0 (a(t_T))^{- 3(\phi_2 - \phi_1)},
\label{rho2}
\end{equation}
so that $(\rho_2)_0 < 0$ and, because $a(t_T) \gg 1$
and if $(\phi_2 - \phi_1)$ is sufficiently non-zero, it follows that
$|(\rho_2)_0| \ll |(\rho_1)_0|$ and hence that the
second ({\it ``$i = 2$"} in Eq.(\ref{Friedman})) dark energy
superfluid is unobservably small at the present time $t=t_0$.

\bigskip

\noindent Can $\rho_2$ play the role of causing both the turnaround and the
bounce in cyclic cosmology? The answer is negative as is now explained by
a no-go theorem.

\bigskip

\noindent {\it No-Go theorem}. ~~
Let us prove a no-go theorem that $n_s = 2$ cannot produce
an acceptable bounce at $t=t_B$ where contraction turns
into expansion.

\bigskip

At the bounce when $t=t_B$, we would need

\begin{equation}
(\rho_2)_0 = - (\rho_r)_0 (a(t_B))^{-(4 + 3\phi_2)}
\label{rho22}
\end{equation}
and, because $a(t_B) \ll 1$, this would require
$\phi_2 < - 4/3$, or $w_2 > + 1/3$, clearly inconsistent
with Eq.(\ref{rho2}) which requires $\phi_2 > 0$ and $w_2 > +1/3$.
This incompatibility of Eqs.(\ref{rho2})
and (\ref{rho22}) provide a No-Go theorem for any four-fluid
($n_n=2$ and $n_s=2$) model.

\bigskip

\noindent This is not surprising when we consider the
brane-world Friedman equation
\cite{BF}
\begin{equation}
\left( \frac{\dot{a}}{a} \right)^2
= \frac{8 \pi G}{3} \left[\rho_{\Lambda} a(t)^{3 \phi_{\Lambda}}
+ \frac{(\rho_m)_0}{a(t)^3}
+ \frac{(\rho_r)_0}{a(t)^4}
- \frac{\rho_{total}^2}{\rho_C} \right],
\label{Baum}
\end{equation}
where $\rho_{total} = (\rho_{\Lambda} +  \rho_m + \rho_r)$. Thus,
the final term on the right-hand-side of Eq.(\ref{Baum}) has
quite a different time dependence
for $t \rightarrow t_T$ and $t \rightarrow t_B$. This
underlies the No-Go theorem and mandates usage of the
following five-fluid model.

\bigskip

\noindent {\it Five-Fluid Model}. ~~
We consider a five-fluid model with
$n_n=2$ and $n_s=3$. Redefine the densities $\rho_2 \rightarrow - \rho_2$
and $\rho_3 \rightarrow - \rho_3$ and the Friedman equation
becomes
\begin{equation}
\left( \frac{\dot{a}}{a} \right)^2
= \frac{8 \pi G}{3} \left[ \frac{(\rho_m)_0}{a(t)^3}
+ \frac{(\rho_r)_0}{a(t)^4}
+ (\rho_1)_0 a(t)^{3\phi_1}
- (\rho_2)_0 a(t)^{3\phi_2}
- (\rho_3)_0 a(t)^{3\phi_3} \right] ,
\label{Friedman5}
\end{equation}
and in this case we can arrange that at the turnaround
\begin{equation}
(\rho_2)_0 = (\rho_1)_0 (a(t_T))^{- 3(\phi_2 - \phi_1)}, 
\label{rho222}
\end{equation}
and at the bounce
\begin{equation}
(\rho_3)_0 = (\rho_r)_0 (a(t_B))^{-(4 + 3\phi_3)}. \label{rho3}
\end{equation}

\bigskip

\noindent At the turnaround, as in the four-fluid model, we require
(I) $(\phi_2 - \phi_1) > 0$ while at the bounce there is the new condition
(II) $(\phi_3 + 4/3) < 0$.

\bigskip

Taking these two inequalities (I) and (II) together will ensure the turnaround
and bounce occur and that $|(\rho_1)_0| > |(\rho_2)_0|$
and $|(\rho_r)_0| > |(\rho_3)_0|$, as necessary to preserve
the successful description of the present universe.
For special values of the equations of state $w_2$ and $w_3$
the five-fluid model becomes indistinguishable from the BF model,
as follows.

\bigskip

\noindent {\it Brane-world as special case of five-fluid model}. ~~
In the special case, consistent with the above inequalities,
where $\phi_2 = 2 \phi_1$ and $\phi_3 = -8/3$ the Friedman
equation of Eq.(\ref{Friedman5}) become indistinguishable
for that of the brane -world model in Eq.(\ref{Baum}).
Although Eq.(\ref{Friedman5}) is not {\it identical} to
Eq.(\ref{Baum}) there is no observable difference because
at present the components $(\rho_2)$ and $(\rho_3)$ are
completely negligible; at turnaround $(\rho_2)$ duplicates the
final term in Eq.(\ref{Baum}) and at the bounce $(\rho_3)$
plays precisely the same role.

\bigskip

\noindent However, the five-fluid model is more general if we incorporate
arbitrary values of $\phi_2$ and $\phi_3$ consistent with the
above inequalities (I) and (II).

\bigskip

\noindent {\it Distinguishing models}. ~~ Let us consider
a five-fluid model which is very disparate from the BF model.
In such a case, accurate observations can distinguish
the cyclic models.

\bigskip

Of course, we do not yet know $\phi_1$ precisely for the dark energy
but let us suppose that $\phi_1=0.05$, consistent with
present WMAP data\cite{WMAP}. We need $\phi_2$ to be bigger so let
us assume $\phi_2=0.06$. In this case Eq.(\ref{rho222})
requires that
\begin{equation}
(\rho_2)_0 = (\rho_1)_0 a(t_T)^{0.03}.
\label{003}
\end{equation}

\bigskip

Just to complete an example, let us now assume $a(t_T) = 10^{33.33}$
whereupon Eq.(\ref{003}) dictates that $(\rho_1)_0 = 10 (\rho_2)_0$
and so the fit to dark energy should use a scale dependence
\begin{equation}
(\rho_{DE})_0 \left[ a(t)^{0.15} - 0.1 a(t)^{0.18} \right].
\label{newfit}
\end{equation}

\bigskip

More generally the multifluid model suggests fitting to
\begin{equation}
(\rho_{DE})_0 \left[ a(t)^{3 \phi_1} - \eta a(t)^{3 \phi_2} \right],
\label{newfitgeneral}
\end{equation}
where $\phi_2 > \phi_1$ and $\eta$ is an additional parameter
related, in general, to the turnaround scale.

\bigskip

More accurate and complete data on dark energy will enable distinction
between fitting with a two-term formula like Eq.(\ref{newfitgeneral}) and
fitting with only the first term.

\bigskip
\bigskip
\bigskip

\noindent {\it Discussion} ~~  Inspired  by previous successes,
we have here attempted to emulate the two-fluid model
of superfluidity in a five-fluid model of cyclic
cosmology. The analogy is heightened by the zero entropy
for the dark energy (superfluid) components,

\bigskip

The multifluid models have certain
advantages, including that they do not necessitate
derivations
from brane worlds in higher dimensionality. The analogies
to the two-fluid model of superfluidity may be posited
directly in four dimensions.

\bigskip

In a certain limit, the five fluid model with two normal
fluids, matter and radiation, and three superfluids for dark
energy become indistinguishable from the model of ref.\cite{BF}.

\bigskip

However, when the five-fluid model become very disparate
from the brane-world model it will be possible to distinguish
them by accurate observations of dark energy as we have
discussed.

\bigskip

It is therefore worth examining, as more and better observational data
become available, whether fits to a two-term expression
as in Eq.(\ref{newfitgeneral}) are more successful for dark energy
than those using only the first term thereof.

\bigskip

In conclusion, cosmological dark energy is perhaps
the most important theory challenge in physics or 
astronomy and has stymied all attempts at its understanding.
Although history does not always repeat itself, it does seem
well worth studying the research works in condensed matter 
theory by {\it e.g.} John Bardeen,
Bogoliubov and Landau. This is precisely what provoked our
present suggestion that dark energy be best regarded as a superfluid, or
superposition of superfluids, of the type very familiar in condensed 
matter. This, in turn,  hints at an alternative to the big bang initial singularity.

\bigskip
\bigskip
\bigskip
\bigskip

\begin{center}

\section*{Acknowledgements}

\end{center}

The work of PHF and SM was supported in part by the U.S. Department of Energy under Grant
No. DG-FG02-06ER41418.
IA was supported in part by the  INTAS grant 03-51-6346.


\bigskip
\bigskip
\bigskip


\begin{thebibliography}{99}

\bibitem{Nambu}
Y. Nambu, Phys. Rev. Lett. {\bf 4,} 380 (1960).
\bibitem{BCS}
J. Bardeen, L.N. Cooper and J.R. Schrieffer, Phys. Rev. {\bf 108,} 1175 (1957).
\bibitem{NNB} N.N.Bogoliubov, J. Phys. (USSR)  {\bf 11} (1947) 23.
\bibitem{BTS} N.N.Bogoliubov, D.V. Shirkov and V.V. Tomachev,
{\it A new method in the theory of
superconductivity}. 
Consultants Bureau (1959).
\bibitem{BF}
L. Baum and P.H. Frampton, Phys. Rev. Lett.
{\bf 98,} 071301 (2007). {\tt hep-th/0610213}.
\bibitem{F}
P.H. Frampton, Mod. Phys. Lett. {\bf A22,} 2587 (2007).  \\
{\tt arXiv:0705.2730}.
\bibitem{BF2}
L. Baum and P.H. Frampton, Mod. Phys. Lett. {\bf A23,} 36 (2008).
{\tt hep-th/0703162}.
\bibitem{Landau}
L.D. Landau, J. Phys. U.S.S.R. {\bf 5,} 71 (1941).
\bibitem{Bardeen}
J. Bardeen, Phys. Rev. {\bf 11,} 399 (1958).
\bibitem{NO}
S. Nojiri and S.D. Odintsov, Phys. Lett. {\bf B637,} 139 (2006). {\tt astro-ph/0603062};
{\it ibid} {\bf B639,} 144 (2006). {\tt hep-th/0606025}.
\bibitem{WMAP}
D.~N.~Spergel {\it et al.}  [WMAP Collaboration],
Astrophys.\ J.\ Suppl.\  {\bf 170}, 377 (2007)
{\tt astro-ph/0603449}.


\end{thebibliography}
\end{document}